\documentclass[twocolumn,preprintnumbers,amsmath,amssymb]{revtex4}
\usepackage{graphicx} % include figure files
\usepackage{setspace} % include figure files
\usepackage{dcolumn} % align table columns on decimal point
\usepackage{bm} % bold math
\usepackage{natbib}
\usepackage{hyperref}
\hypersetup{colorlinks=true,linkcolor=blue,citecolor=blue,urlcolor=blue}
\usepackage{epsfig}
\usepackage{amsmath}
\begin{document}
\title{Compositionally Engineered Non-Equimolar LaCoO$_3$-Based High-Entropy Perovskites with Enhanced Thermoelectric Performance}

\author{Jitendra Kumar$^a$}
\author{David Bérardan$^b$}
\author{Diana Dragoe$^b$}
\author{Nita Dragoe$^b$}
\author{Ashutosh Kumar$^{a,}$\footnote{Email: ashutosh@iitbhilai.ac.in}}
\affiliation{$^a$Functional Materials Laboratory, Department of Materials Science and Metallurgical Engineering, Indian Institute of Technology Bhilai, Chhattisgarh 491002, India}
\affiliation{$^b$ICMMO (UMR CNRS 8182), Université Paris-Saclay, F-91405 Orsay, France}

\date{\today}
\begin{abstract}

The decoupling of phonon and electron transport remains a central challenge in the development of high-performance thermoelectric materials. Configurational-entropy maximization is widely invoked as the design principle for decoupling phonon and electron transport. The present study investigates a compositionally engineered LaCoO$_3$-based high-entropy perovskites to determine whether thermoelectric transport can be improved by tuning cation identity and concentration rather than maximizing configurational entropy. La$_{1-x}$Sr$_x$(CoFeMnCrNi)O$_3$ ($0 \leq x \leq 0.2$) and selected non-equimolar A- and B-site perovskite compositions were prepared by solid-state reaction. The obtained samples are predominantly single-phase, as confirmed by XRD and Rietveld refinement, in agreement with the calculated size-disorder parameters, while multication disorder provides strong mass- and strain-field fluctuations for phonon scattering. All samples exhibit p-type, thermally activated transport, and is found to be consistent with adiabatic small-polaron hopping. Sr substitution progressively reduces the hopping barrier and electrical resistivity, whereas non-equimolar B-site engineering partially recovers electrical transport while retaining low thermal conductivity. La$_{0.9}$Sr$_{0.1}$Co$_{0.4}$Cr$_{0.3}$Ni$_{0.1}$Fe$_{0.1}$Mn$_{0.1}$O$_3$, which combines a coupled Co-rich and Cr-rich B-site, providing mixed valence and spin-state degeneracy that sustain a large thermopower from Co, while Cr limits carrier concentration while lowering the polaron hopping barrier to 0.19 eV, delivering a power factor of 40–43 $\mu$W·m$^{-1}$·K$^{-2}$ and zT $\sim$ 0.072 at 1100 K, 2.7 times higher the equimolar analogue. The results demonstrate that cation chemistry and mass contrast, rather than configurational-entropy maximization alone, provide a more effective strategy for balancing electronic and phonon transport in multicomponent oxide thermoelectric.\\
\end{abstract}
\maketitle
\section{Introduction}
The quest for cleaner and more efficient energy technologies has become one of the most pressing scientific and technological challenges of our time, driven by the need to mitigate climate change and reduce dependence on finite fossil-fuel resources. In this context, thermoelectric (TE) materials, which can directly convert waste heat into useful electrical energy and vice versa, have attracted considerable attention for sustainable energy conversion and solid-state cooling 
applications~\cite{rowe2018crc, terasaki2005introduction}. Oxide TE materials are especially attractive because of their excellent chemical and thermal stability at elevated temperatures in air, low toxicity, and relatively low synthesis cost compared with conventional chalcogenide- and pnictide-based systems~\cite{he2011oxide}. The thermoelectric performance of a material is evaluated by the dimensionless 
figure of merit,
\[
zT = \frac{\alpha^2 \sigma T}{\kappa},
\]
where $\alpha$ is the Seebeck coefficient, $\sigma$ is the electrical conductivity, $\kappa$ is the total thermal conductivity, and $T$ is the absolute temperature. The total thermal conductivity generally consists of electronic and phononic contributions, i.e., $\kappa = \kappa_{\mathrm{e}} + \kappa_{\mathrm{ph}}.$\\
In oxide materials, $\kappa_{\mathrm{ph}}$ usually dominates the total thermal conductivity, while low $\sigma$ further limits the overall TE performance. In other words, despite their chemical and thermal robustness, the practical deployment of oxide TE is still hindered by their relatively poor TE performance, mainly arising from high $\kappa_{\mathrm{ph}}$ and/or poor $\sigma$. Therefore, developing new design strategies capable of simultaneously suppressing phonon transport while maintaining or improving charge transport remains a central objective in oxide TE research~\cite{koumoto2010oxide,wang2026advances}.\\
Among emerging approaches for oxide TE, multicomponent cation substitution has gained significant momentum as a means of suppressing $\kappa_{\mathrm{ph}}$. Such systems are conventionally classified by the configurational entropy of mixing, $\Delta S_{\mathrm{conf}}$ = -R $\sum_i x_i \ln x_i$, where $x_i$ is the atomic fraction of the $i$$^{th}$ species and since the first report of entropy-stabilized oxides in 2015~\cite{rost2015entropy}, the field has largely organized itself around maximizing $\Delta S_{\mathrm{conf}}$, with five or more cations placed on a crystallographic site in near-equimolar proportions. It is important to recognize, however, what $\Delta S_{\mathrm{conf}}$ does? It is a purely combinatorial quantity determined by the number of species and their fractions; it is invariant under exchange of one cation for another of equal fraction. The physical mechanisms that actually govern transport-point-defect phonon scattering, which scales with mass and radius contrast; carrier concentration, which is set by the accessible valence states; and the entropy transported per carrier, which is set by spin and orbital degeneracy — are all chemistry-specific and are therefore not captured by $\Delta S_{\mathrm{conf}}$~\cite{mccormack2021thermodynamics, banerjee2020high}.\\
Multicomponent cation substitution has been shown to suppress $\kappa$ across a wide range of oxide structure families. In perovskite systems (ABO$_3$), multication occupancy of either the A-site or the B-site substantially reduces $\kappa$. For example, the A-site high-entropy perovskite (Ca$_{0.2}$Sr$_{0.2}$Ba$_{0.2}$Pb$_{0.2}$La$_{0.2}$)TiO$_3$ exhibits 
$\kappa_{\mathrm{ph}}$ of 1.09~W/(m·K) at 923~K and achieves $zT \sim 0.2$ at 873~K after reduction treatment~\cite{zhang2022high}. Likewise, the B-site multi-component perovskite 
Sr(Ti$_{0.2}$Fe$_{0.2}$Mo$_{0.2}$Nb$_{0.2}$Cr$_{0.2}$)O$_3$ shows an ultralow $\kappa$ of $\sim$0.7~W/(m·K) at 1100~K, highlighting the remarkable ability of multication disorder to impede phonon transport~\cite{banerjee2020high}. Sr$_{0.9}$La$_{0.1}$(Zr$_{0.25}$Sn$_{0.25}$Ti$_{0.25}$Hf$_{0.25}$)O$_3$ exhibits low $\kappa$ together with high $\alpha$, demonstrating the role of high entropy in enhancing the overall thermoelectric response of SrTiO$_3$-derived oxides~\cite{lou2022novel}, while an A-site-deficient high-entropy perovskite achieves $zT \sim 0.24$ at 1073~K~\cite{zhang2023thermoelectric}. These examples illustrate that multicomponent substitution provides a versatile platform for reducing $\kappa$.\\
The beneficial role of multicomponent substitution is not limited to perovskites. In tungsten-bronze systems, rare-earth-free high-entropy (Sr$_{0.2}$Ba$_{0.2}$Li$_{0.2}$K$_{0.2}$Na$_{0.2}$)Nb$_2$O$_6$ delivers a large $\alpha$ and a very low $\kappa \sim 0.8$~W/(m·K) at 330\,K, 
leading to a peak $zT$ of 0.23 at 1150~K~\cite{jana2023designing}. 
(Na$_{0.1}$K$_{0.1}$Sr$_{0.2}$Ba$_{0.2}$La$_{0.2}$Bi$_{0.2}$)Nb$_2$O$_{6-\delta}$ shows $zT = 0.19$ at 1023~K, confirming that entropy engineering is also effective in niobate-based oxide systems~\cite{zhu2025thermoelectric}. In fluorite-structured oxides, Mg--Zn co-doped high-entropy ceramics based on 
(Zr$_{0.194}$Ce$_{0.194}$Pr$_{0.194}$Y$_{0.194}$Ho$_{0.194}$Mg$_x$Zn$_{0.03-x}$)O$_{2-\delta}$ 
have shown one of the best performances among high-entropy oxide thermoelectrics, reaching $zT = 0.46$ at 773~K through synergistic optimization of charge and heat transport~\cite{Ji2026synergistic}. A significant reduction in $\kappa \sim$ 2.1 W/(m·K) is achieved in high-entropy wolframite oxide compared with MgWO$_4$ ($\kappa \sim$ 11.5 W/(m·K))~\cite{kumar2024thermoelectric}. In Bi$_2$Sr$_2$Co$_2$O$_y$-based systems, high-entropy substitution combined with Cu/Ag doping further improves TE performance, reaching $zT \sim 0.36$ at 923~K~\cite{leng2026enhancement}. These studies 
collectively establish that configurational entropy provides a versatile platform for suppressing $\kappa$ and, when judiciously combined with texturing, doping, or compositing, can also enhance electrical transport.\\
Although large configurational entropy is highly effective in suppressing $\kappa_{\mathrm{ph}}$, excessive lattice disorder can also localize carriers and reduce mobility, thereby limiting the achievable power factor. Therefore, the design of high-entropy oxide thermoelectrics must move beyond simply maximizing configurational entropy and instead target a balanced optimization of 
both phonon and charge transport. Cobaltate-based perovskites are particularly promising owing to their favorable p-type conductivity, relatively large Seebeck coefficient, mixed Co$^{3+}$/Co$^{4+}$ valence states, and spin-state-dependent electronic structure arising from strong Co $3d$--O $2p$ hybridization, which collectively govern charge transport and thermoelectric behavior. Motivated by our earlier work on A-site high-entropy rare-earth cobaltates, (LaNdPrSmEu)$_{1-x}$Sr$_x$CoO$_3$, which showed that introducing multiple rare-earth ions at the A-site significantly reduces $\kappa_{\mathrm{ph}}$ and improves $\alpha$, leading to a maximum $zT$ of 0.23 at 350~K for (LaNdPrSmEu)$_{0.95}$Sr$_{0.05}$CoO$_3$~\cite{kumar2023thermoelectric}, the present study explores B-site and dual-sublattice compositional engineering in rare-earth cobaltate 
perovskites to identify a disorder regime that minimizes thermal conductivity without excessively compromising electronic transport.\\
In this study, we first examine the La$_{1-x}$Sr$_x$(CoFeMnCrNi)O$_3$ ($0 \leq x \leq 0.2$) series to understand the effect of Sr substitution in an equimolar high-entropy B-site configuration. We then extended the compositional design to non-equimolar B-site distributions and selected multi-rare-earth A-site compositions to tune carrier transport without sacrificing the thermal-transport advantages of chemical complexity. Unlike most high-entropy thermoelectric studies that emphasize maximizing configurational entropy to lower lattice thermal conductivity, the present work demonstrates that excessive B-site disorder can penalize electronic transport and power factor. Therefore, a non-equimolar chemical species and their concentration is adopted to identify an optimal disorder regime in which phonon scattering remains strong while electronic transport is partially recovered.\\
\section{Experimental Details}
High-entropy rare-earth perovskite oxides La$_{1-x}$Sr$_x$(CoFeMnCrNi)O$_3$ ($0 \leq x \leq 0.2$): $x = 0.00$ (T03a), $x = 0.05$ (T03b), $x = 0.10$ (T03c), and $x = 0.20$ (T03d); T03f: 
La$_{0.9}$Sr$_{0.1}$(Co$_{0.15}$Fe$_{0.15}$Mn$_{0.15}$Cr$_{0.15}$Ni$_{0.4}$)O$_3$; 
T03g: La$_{0.9}$Sr$_{0.1}$(Co$_{0.3}$Ni$_{0.2}$Fe$_{0.3}$Mn$_{0.1}$Cr$_{0.1}$)O$_3$; 
T03j: La$_{0.9}$Sr$_{0.1}$(Co$_{0.4}$Cr$_{0.3}$Ni$_{0.1}$Fe$_{0.1}$Mn$_{0.1}$)O$_3$; 
T03k: (LaNdPrDySm)$_{0.9}$Sr$_{0.1}$(Co$_{0.3}$Ni$_{0.2}$Fe$_{0.3}$Mn$_{0.1}$Cr$_{0.1}$)O$_3$; 
and T03m: (LaNdPrDySm)$_{0.9}$Sr$_{0.1}$ (Co$_{0.4}$Cr$_{0.3}$Ni$_{0.1}$Fe$_{0.1}$Mn$_{0.1}$)O$_3$ 
were synthesized using a standard solid-state reaction route. Stoichiometric amounts of La$_2$O$_3$ (Alfa Aesar, 99.9\%), Nd$_2$O$_3$ (Alfa Aesar, 99.9\%), Pr$_6$O$_{11}$ (Alfa Aesar, 99.9\%), Dy$_2$O$_3$ (Alfa Aesar, 99.9\%), Sm$_2$O$_3$ (Alfa Aesar, 99.9\%), SrCO$_3$ (Alfa Aesar, 99.9\%), Co$_3$O$_4$ (Alfa Aesar, 99.9\%), MnO$_2$ (Alfa Aesar, 99.9\%), Cr$_2$O$_3$ (Alfa Aesar, 99.9\%), NiO (Alfa Aesar, 99.9\%) and Fe$_2$O$_3$ (Alfa Aesar, 99.9\%) were mixed using planetary ball milling. The mixture was then pressed into pellets and sintered at 1250--1300$^\circ$C for 48~h with intermediate grinding, using heating and cooling rates of 200$^\circ$C·h$^{-1}$. Phase formation and crystal structure were examined by powder X-ray diffraction using Cu-K$\alpha$ radiation ($\lambda = 1.5406$~\AA) on a Bruker D8 diffractometer, followed by Rietveld refinement~\cite{rietveld1969profile}. The sintered pellets were cut into bar-shaped specimens for electrical resistivity and Seebeck coefficient measurements. Microstructural analysis was performed using field-emission scanning electron microscopy (FE-SEM, Zeiss Sigma HD), whereas high-resolution transmission electron microscopy (HRTEM, JEOL JEM-F200 (CF-HR)) equipped with energy-dispersive X-ray spectroscopy (EDS) mapping was used to probe the local structure and elemental homogeneity. For HRTEM measurements, the powder sample was dispersed in ethanol medium using
ultrasonication for 4 hours followed by adding one drop of suspension on copper grid, which was dried overnight in
a desiccator. The surface chemical state was analyzed using X-ray photoelectron spectroscopy (XPS) with a Thermo
$\mathrm{K}_{\alpha}^{+}$ spectrometer, with a monochromatic Al $\mathrm{K}\alpha$ X-ray source (1486.7 eV) operating at 6 mA and 12 kV (72W), and an X-ray spot size of $400\,\mu\mathrm{m}$, corresponding to an irradiated area of about $1\,\mathrm{mm}^{2}$. The base pressure was $2 \times 10^{-9}$ mbar. The angle analyser/X-Ray source is 56°. Charge compensation was achieved using a dual beam flood gun using low energy electrons $(<5\,\mathrm{eV})$ in a partial argon atmosphere $(2.4 \times 10^{-8}\,\mathrm{mbar})$. The survey scans were
measured at a pass energy of 200 eV with a step size of 1 eV, using a dwell time of 10 ms, while narrow scans were
measured at a pass energy of 50 eV with a step size of 0.1 eV and 50 ms dwell time. The binding energy referencing
was done with respect to the C-C chemical environment in the C 1s photopeak, set at 284.8 eV, which gives the
oxide component of O1s photopeak at 529.4\,eV. The precision in binding energy is ± 0.2 eV. The core-level
spectra were treated using the Avantage software provided by the manufacturer, by implementing the RSF from the
Thermo Fisher library, after the subtraction of a Shirley-type background. The core-level spectra were fitted using a
Shirley background with mixed Gaussian–Lorentzian line shapes, with 30\% Lorentzian character. Electrical resistivity and Seebeck coefficient were measured using a home-built setup from 300 to 1100~K.\cite{byl2012experimental} Thermal conductivity was calculated using $\kappa = D·d·C_p,$ where $D$ is the thermal diffusivity measured from 300 to 1100~K using a Netzsch LFA-457 laser flash analyzer, $d$ is the bulk density obtained from the sample mass and geometric volume, and $C_p$ was estimated using the Dulong--Petit approximation. The experimental uncertainties in $\rho$, $\sigma$ and $\kappa$ are 6\%, 6\% and 5\% respectively. The values of relative density for these samples are found to be in the range of 75-81, as shown in Table~\ref{tab:structural_parameters}.\\
\begin{table*}[htbp]
\centering
\caption{Sample code, composition, tolerance factor ($t$), B-site size-disorder parameter ($\delta$), B- and A-site mass-variance parameters ($\Gamma_M$), and configurational entropy ($\Delta S_{\mathrm{conf}}$) for all compositions.}
\label{tab:structural_parameters}
\begin{tabular}{cccccccc}
\hline
Sample & Composition &$t$ & $\delta$ & $\Gamma_M(B)$ & $\Gamma_M(A)$ & $\Delta S_{\mathrm{conf}}$& Rel. Density\\
 &  & & (\%) & $10^{-3}$ & $10^{-3}$ & $(R)$ & (\%) \\
\hline
T03a & La(CoFeMnCrNi)O$_3$&0.956 & 4.45 & 2.10& 0.00&1.61 & 79\\
T03b & La$_{0.95}$Sr$_{0.05}$(CoFeMnCrNi)O$_3$&0.958 & 4.45 & 2.10 & 6.72 & 1.81 & 75\\
T03c &La$_{0.9}$Sr$_{0.1}$(CoFeMnCrNi)O$_3$ &0.959 & 4.45 & 2.10 & 13.23 & 1.93 & 78\\
T03d & La$_{0.8}$Sr$_{0.2}$(CoFeMnCrNi)O$_3$&0.962 & 4.45 & 2.10 & 25.43 & 2.11 & 81\\
T03f & La$_{0.9}$Sr$_{0.1}$(Co$_{0.15}$Fe$_{0.15}$Mn$_{0.15}$Cr$_{0.15}$Ni$_{0.4}$)O$_3$&0.953 & 5.42 & 1.94& 13.23& 1.83 & 79\\
T03g &La$_{0.9}$Sr$_{0.1}$(Co$_{0.3}$Ni$_{0.2}$Fe$_{0.3}$Mn$_{0.1}$Cr$_{0.1}$)O$_3$ &0.959 & 4.52 &1.55& 13.23& 1.83 & 78\\
T03j & La$_{0.9}$Sr$_{0.1}$(Co$_{0.4}$Cr$_{0.3}$Ni$_{0.1}$Fe$_{0.1}$Mn$_{0.1}$)O$_3$&0.966 & 4.49 &2.88& 13.23& 1.74 & 81\\
T03k & (LaNdPrDySm)$_{0.9}$Sr$_{0.1}$(Co$_{0.3}$Ni$_{0.2}$Fe$_{0.3}$Mn$_{0.1}$Cr$_{0.1}$)O$_3$&0.918 & 4.52 &1.55& 19.33& 3.28& 76\\
T03m & (LaNdPrDySm)$_{0.9}$Sr$_{0.1}$(Co$_{0.4}$Cr$_{0.3}$Ni$_{0.1}$Fe$_{0.1}$Mn$_{0.1}$)O$_3$ &0.924 & 4.50 & 2.88& 19.33&3.19&79\\
\hline
\end{tabular}
\end{table*}
%
%
%
%\subsection{Computational Methods}
%The Density Functional Theory (DFT)\cite{kresse1996efficiency,kohn1965self} calculations were performed using the Vienna \textit{Ab}-\textit{initio} Simulation Package (VASP)\cite{kresse1996efficiency,kresse1999ultrasoft}. The Projector-Augmented Wave (PAW)\cite{kresse1999ultrasoft,blochl1994projector} pseudopotential was used to model the interaction between the ion cores and the valence electrons, with a cutoff energy of 520 eV. In order to accurately capture correlation effects of 3d electrons in transition metals, we adopted the spin-polarized local density approximation combined with the Hubbard U correction (LSDA + U) \cite{dudarev1998electron, liechtenstein1995density}. The Perdew-Zunger parametrization \cite{perdew1981self} of the Ceperly-Alder model \cite{ceperley1980ground} was chosen for the exchange-correlation part of LSDA. To incorporate the Hubbard U correction, we applied the Dudarev formalism, setting U$_{eff}$ to 3.32 eV, 3.7 eV, 3.9 eV, 5.3 eV, and 6.2 eV for Co, Cr, Mn, Fe, and Ni, respectively \cite{zhou2004first,cococcioni2005linear,wang2004enthalpy,wang2006oxidation,jain2011formation}. All the structures were fully optimized (atomic positions and lattice parameters) until the Hellmann-Feynman forces were less than 10$^{-2}$ eV/\AA. The tolerance for the electronic self-consistent iteration was kept at 10$^{-5}$ eV. A $\Gamma$-centered $\mathit{k}$-point mesh of 2 $\times$ 4 $\times$ 4 was used for the Brillouin zone sampling.
%
%
%
\section{Results and Discussion} 
\subsection{Equimolar B-site high entropy samples: La$_{1-x}$Sr$_x$(CoFeMnCrNi)O$_3$}
\subsubsection{X-ray diffraction}
\begin{figure*}
\centering
  \includegraphics[width=0.9\linewidth]{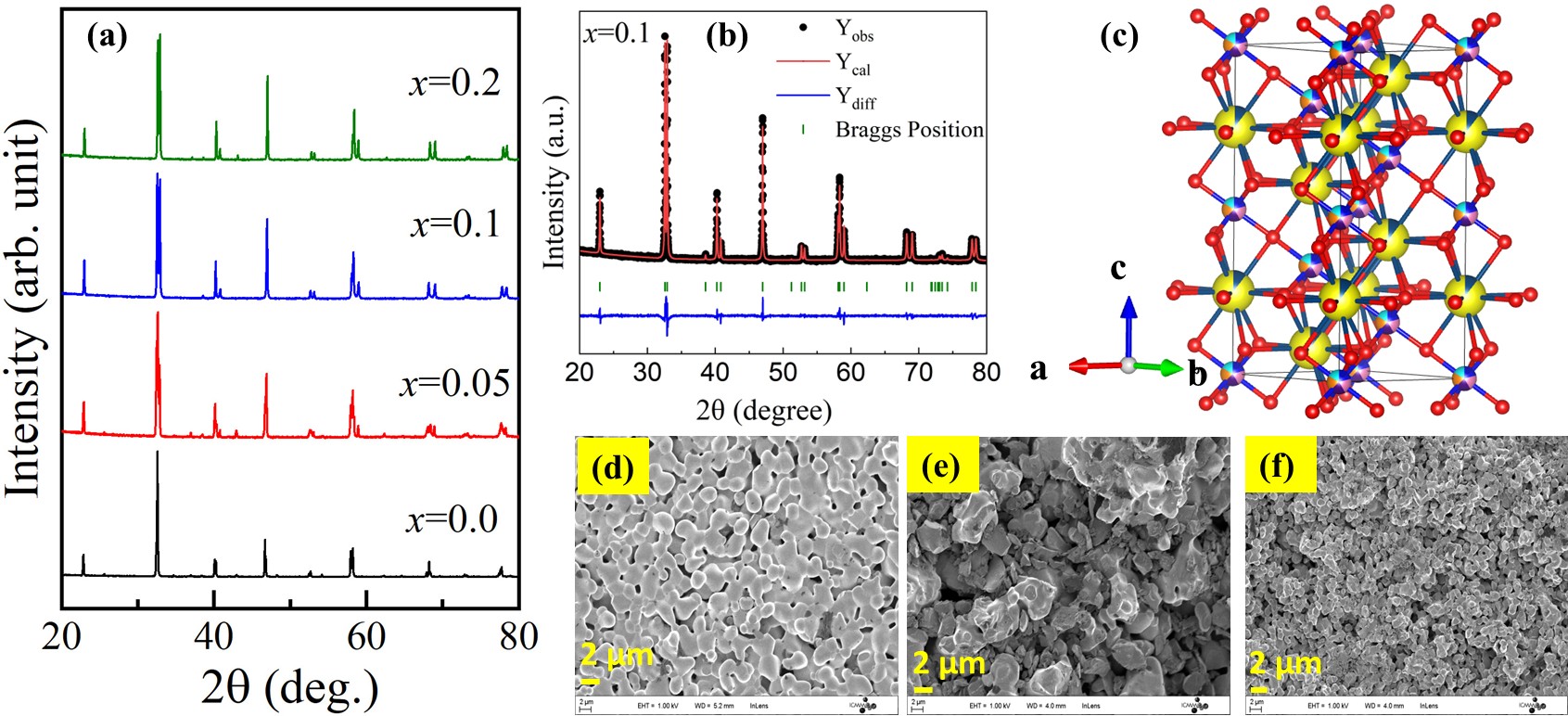}
  \caption{(a) X-ray diffraction (XRD) patterns of La$_{1-x}$Sr$_x$(CoFeMnCrNi)O$_3$ ($0 \leq x \leq 0.2$); (b) Rietveld refinement pattern for the $x = 0.1$ composition; (c) unit cell drawn using the structural parameters obtained from Rietveld refinement using VESTA software; and field-emission scanning electron microscopy (FESEM) images of La$_{1-x}$Sr$_x$(CoFeMnCrNi)O$_3$ pellets for (d) $x = 0.00$, (e) $x = 0.05$, and (f) $x = 0.10$.}
  \label{fig1}
\end{figure*}
The XRD patterns of $\mathrm{La}_{1-x}\mathrm{Sr}_{x}(\mathrm{CoFeMnCrNi})\mathrm{O}_{3}\quad(0\leq x\leq 0.2)$ is shown in Fig.~\ref{fig1}(a). The reflections corresponding to secondary phases are negligible in all samples (well below 1\%) indicating that all five B-site cations are successfully incorporated into the perovskite lattice. We note that phase stability here is adequately rationalised by the tolerance factor and by $\delta \leq$5\%. The tolerance factor ($t$) for all the samples synthesized in the present study was calculated using $t = \frac{r_A + r_O}{\sqrt{2}(r_B + r_O)},$ where $r_A$ and $r_B$ are the average ionic radii at the A- and B-sites, respectively, and $r_O$ is the ionic radius of oxygen. (A-site elements were taken to +3 ($\mathrm{Sr}^{2+}$) be in XII/IX coordination and B-site elements were considered
in +3 charge states and VI coordination). The value of $t$ is found to be 0.956 for T03a and increases slightly to 0.962 for T03d (Table~\ref{tab:structural_parameters}), indicating that all compositions retain a similar crystal structure, as confirmed by XRD. Furthermore, the size-disorder parameter $\delta$ (\%) was calculated using the following expression:
\begin{equation}
\delta(\%) = 100 \sqrt{
\sum_{i=1}^{n} c_i
\left(
1 - \frac{r_i}{\sum_{j=1}^{n} c_j r_j}
\right)^2
}.
\end{equation}
Here, $\sum_{j=1}^{n} c_j r_j$ represents the average atomic radius, while $r_i$ and $c_i$ are the ionic radius and atomic fraction of the $i$th element, respectively. The calculated values are listed in Table~\ref{tab:structural_parameters}. The size-disorder parameter for La$_{1-x}$Sr$_x$(CrMnFeCoNi)O$_3$ ($0 \leq x \leq 0.2$) lies in the range 0 $\leq \delta \leq$5, which is considered favorable for the formation of a single-phase sample~\cite{zhang2008solid}. In addition to the size-disorder parameter, the mass contrast of each cation sublattice was quantified using the Abeles mass-variance parameter,
\begin{equation}
\Gamma_{M} \;=\; \sum_{i} c_{i}\left(1-\frac{M_{i}}{\bar{M}}\right)^{2},
\qquad
\bar{M} \;=\; \sum_{i} c_{i} M_{i},
\label{eq:gamma_M}
\end{equation}
where $M_{i}$ and $c_{i}$ are the atomic mass and atomic fraction of the $i$-th cation on the sublattice considered. Unlike $\Delta S_{\mathrm{conf}}$, $\Gamma_{M}$ is sensitive to which elements are present and is the quantity that enters the Klemens--Callaway description of point-defect phonon scattering. Values of $\Gamma_{M}$ for the A- and B-sublattices of all
compositions are listed in Table~\ref{tab:structural_parameters}. The sample with \textit{x}=0.00 and 0.05 crystallizes in the orthorhombic \textit{Pnma}
structure with slight increase in the lattice parameter from undoped (\textit{x}=0.00) to \textit{x}=0.05 sample. With
further increase in the Sr substitution, the structure transforms from orthorhombic \textit{Pnma} to rhombohedral
\textit{R$\bar{3}$c}, indicating a reduction in the degree of octahedral tilting. The lattice parameter increases slightly from
\textit{x}=0.1 to \textit{x}=0.2. This expansion within the \textit{R$\bar{3}$c} phase is consistent with the larger ionic radii of $\mathrm{Sr}^{2+}$
compared to $\mathrm{La}^{3+}$. The lattice parameters obtained from these samples after Rietveld refinement are shown in Table~\ref{tab:refinement} and the
refinement patterns are shown in Fig.~S1 (Supplementary Information). There are small impurity peaks observed in \textit{x}=0.2 (at $2\theta = 37.2^\circ$ and $43.3^\circ$, which may correspond to some unreacted NiO phase). The refinement pattern for $x = 0.1$ is shown in Fig.~\ref{fig1}(b). The unit cell was drawn using VESTA software employing the refinement parameters obtained for $x = 0.1$ from the Rietveld refinement, as shown in Fig.~\ref{fig1}(c).\\
\begin{figure*}
\centering
  \includegraphics[width=0.85\linewidth]{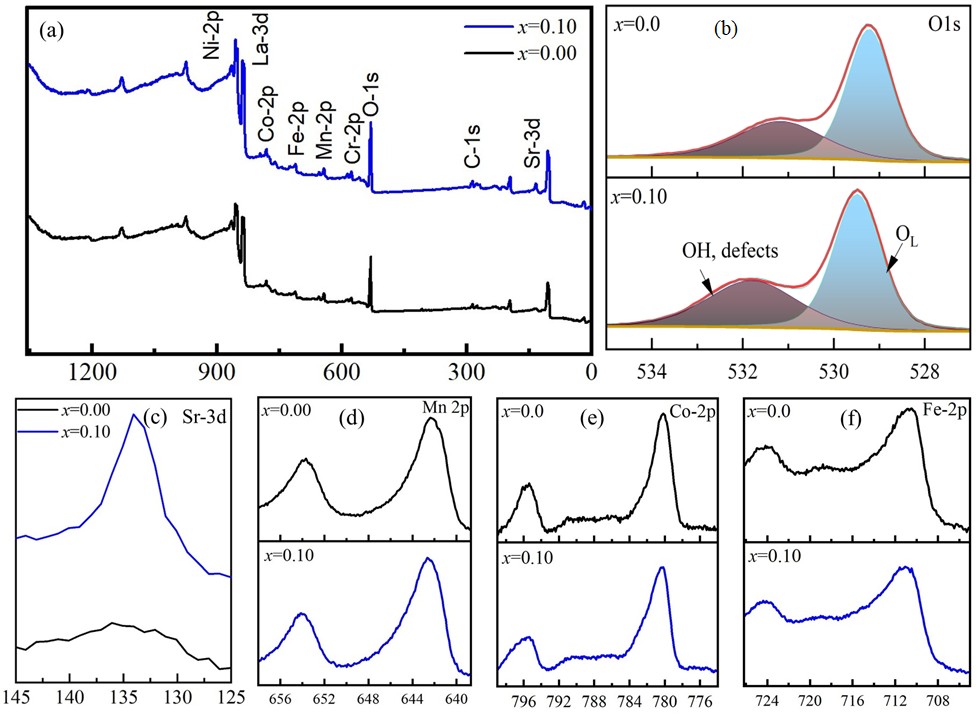}
  \caption{X-ray photoelectron spectroscopy (XPS) spectra of La$_{1-x}$Sr$_x$(CoFeMnCrNi)O$_3$ ($0 \leq x \leq 0.2$): (a) full survey spectrum; core-level spectra of (b) O-1$s$, (c) Sr-3$d$ (d) Mn-2$p$, (e) Co-2$p$, and (f) Fe-2$p$.}
  \label{fig2}
\end{figure*}
FESEM micrographs, shown in Fig.~\ref{fig1}(d--f), reveal the development of polycrystalline microstructures. The microstructure consisting of interconnected grains with some residual porosity. For the un-doped sample (\textit{x}=0.00, Fig.~\ref{fig1}(d)), the microstructure consists of relatively uniform, rounded particles that are closely connected to one another. The particles form an open and interconnected porous network, with irregular voids distributed throughout the surface. For \textit{x}=0.05 (Fig~\ref{fig1}(e)), the morphology changes considerably. The microstructure becomes more heterogeneous, with larger irregular agglomerates, rough angular region and fewer clearly resolved particles. The particle size distribution appears to be broader as compared to that of undoped sample. This coarsening and agglomeration may be associated with the Sr doping which may modify defect concentration, surface energy and diffusion processes during synthesis. With further increase in Sr doping (\textit{x}=0.1), the morphology becomes considerably finer and more
homogeneous. The surface is composed of densely packed, relatively small particles with a larger number of fine interparticles pores. SEM image for \textit{x}=0.2, shown in Fig.~S4, shows no obvious large secondary-phase regions within the examined areas, consistent with the predominantly single-phase perovskite structure identified by XRD. The observed porosity is also consistent with the measured relative densities of approximately 75–81\%.\\
\begin{figure*}
\centering
  \includegraphics[width=0.8\linewidth]{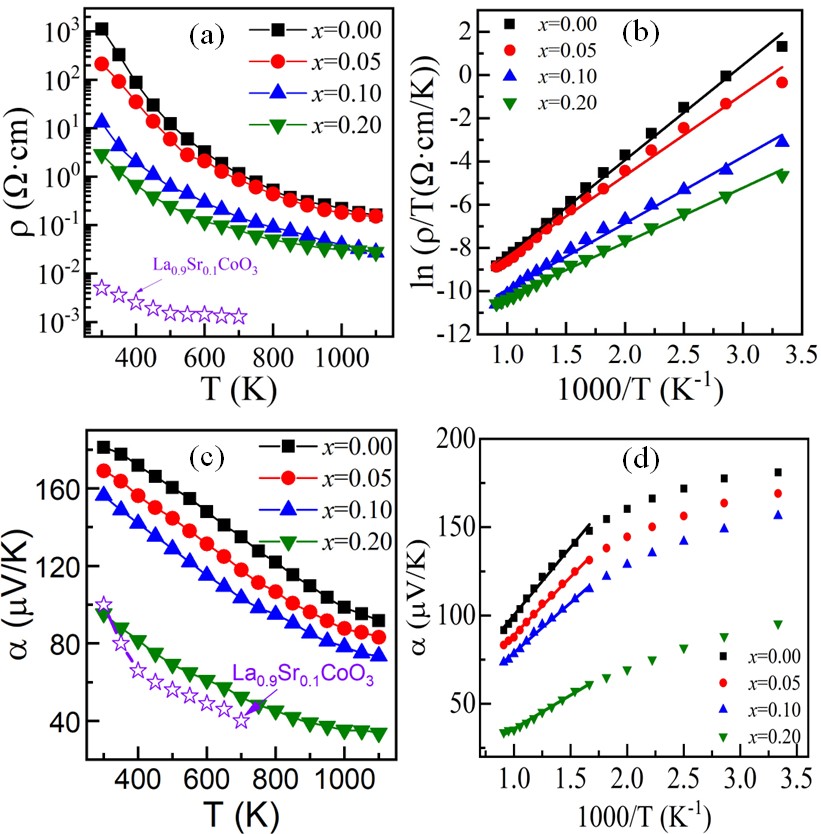}
  \caption{Temperature dependence of (a) electrical resistivity ($\rho$), (b) ln$\rho$ vs 1000/T, (c) Seebeck coefficient ($\alpha$), and (d) $\alpha$ vs 1000/T depicting small polaron hopping model for La$_{1-x}$Sr$_x$(CoFeMnCrNi)O$_3$ ($0 \leq x \leq 0.2$).}
  \label{fig3}
\end{figure*}
\subsubsection{X-ray photoelectron Spectroscopy}
X-ray photoelectron spectroscopy was used to probe the surface chemical states and the oxygen-defect environment of La$_{1-x}$Sr$_x$(CoFeMnCrNi)O$_3$ for $x = 0.00$ and $x = 0.10$. All binding energies were referenced to the adventitious C-1$s$ line at 284.8~eV, which places the oxide component of O-1s spectrum at 529.4 eV and the core-level spectra were fitted using a Shirley background with mixed Gaussian--Lorentzian line shapes, with 30\% Lorentzian character. The survey spectra, as shown in Fig.~\ref{fig2}(a), confirm the presence of all constituent elements, with clearly resolved La-3$d$, Ni-2$p$, Co-2$p$, Fe-2$p$, Mn-2$p$, Cr-2$p$, O-1$s$ and C-1$s$ signals, together with the Sr-3$d$ feature that emerges for $x = 0.10$. The O-1s core-level spectra were fitteed with two components: lower binding energy component at 529.2$\pm$0.2\,eV corresponds to lattice oxygen (O$_L$) of the perovskite framework, wehreas the component centred at 531.5\,eV (lebelled as O-H defects) arises from surface hydroxyl groups, absorbed species and defects. The realtive area of the defect-related component increases from 0.53 for \textit{x}=0.00 to 0.64 for \textit{x}=0.10, indicating that Sr substitution promotes the formation of oxygen-deficient surface enviroments.\\
Since Sr$^{2+}$ substitution at the La$^{3+}$ site requires charge compensation either by oxidation of the B-site cations or by oxygen-vacancy formation, this increase is consistent with the latter mechanism contributing appreciably. Fig.~\ref{fig2}(c) shows the Sr-3$d$ region with no resolvable doublet for $x = 0.00$, whereas a well-defined peak centred near 133~eV appears for $x = 0.10$, confirming the successful incorporation of Sr at the A-site~\cite{young1985xps}. The Mn 2$p$ spectra (Fig.~\ref{fig2}(d)) exhibit the spin--orbit split Mn 2$p_{3/2}$ and Mn 2$p_{1/2}$ components near 642 and 653.1~eV, with a separation of approximately 11.1~eV~\cite{chastain1992handbook}. The position of the 2$p_{3/2}$ line is higher than that expected for Mn$^{2+}$ (640.5--641~eV) and also the absence of the satellite indicate that manganese is present predominantly as a mixture of Mn$^{3+}$ and Mn$^{4+}$. The Co-2$p$ spectra consist of the spin--orbit split Co 2$p_{3/2}$ ($\sim$780~eV) and Co 2$p_{1/2}$ ($\sim$795.5~eV) components with a separation of approximately 15.5~eV (Fig.~\ref{fig2}(e)). The satellite intensity in the 785--790~eV region remains weak, in contrast to the intense satellites characteristic of high-spin Co$^{2+}$; both observations are consistent with cobalt being present predominantly as low-spin Co$^{3+}$, as expected for a LaCoO$_3$-derived lattice
\cite{biesinger2011resolving}. A small fraction of Co$^{4+}$ is expected on Sr substitution, but its presence cannot be confirmed or quantified because the Co-2$p$ region is overlapped by Ni and Fe Auger features. The Fe-2$p$ spectra, shown in Fig.~\ref{fig2}(f), display the Fe-2$p_{3/2}$ line near 711~eV and the Fe 2$p_{1/2}$ line near 724.5~eV, together with a broad shake-up satellite in the 718--720~eV region. Both the 2$p_{3/2}$ binding energy and the position of the satellite are characteristic of Fe$^{3+}$ in an octahedral oxide environment, as found in LaFeO$_3$ and related rare-earth ferrites~\cite{wang20092p3}. Owing to the multicomponent nature of the system, several core levels overlap (La-3$d$ with Ni 2$p$; Co-2$p$ and Fe-2$p$ with Ni and Fe Auger features; Mn-2$p$ with a Ni Auger signal), which prevents reliable quantification of the individual transition-metal valence states. The Cr-2$p$ region overlaps strongly and is therefore not analysed here. The oxygen vacancy presence may not be proven by XPS. This technique may show some oxidataion of the surface (manganese) and can not completely describe the sample surface because of the interferences but allows us to confirm the oxidation state of the elements of interest.\\
\subsubsection{Seebeck Coefficient, Electrical Resistivity, Thermal Conductivity and Figure of Merit:}
Fig.~\ref{fig3}(a) shows the temperature variation of the resistivity for La$_{1-x}$Sr$_x$(CoFeMnCrNi)O$_3$. The resistivity decreases with increasing temperature, confirming non-degenerate semiconducting transport behavior. At 300~K, the resistivity is 1.15~k$\Omega\cdot$cm for $x = 0.0$. The resistivity values for parent LaCoO$_3$ is ~80 $\Omega\cdot$cm\cite{kumar2018improvement}, LaFeO$_3$ is~4.15 $\Omega\cdot$cm\cite{karthikeyan2018thermoelectric}, LaNiO$_3$ is 0.001 $\Omega\cdot$cm\cite{xu1993resisitivity}, LaMnO$_3$ is $\sim$300 $\Omega\cdot$cm\cite{khan2011small} and $\mathrm{LaCrO}_{3}$ is $47\,\mathrm{k}\Omega\cdot\mathrm{cm}$ \cite{jin1994effects}. This shows that the resistivity values obtained for the equimolar high-entropy sample is quite higher than their parent compounds. Further, the resistivity of high entropy sample decreases to 0.2~k$\Omega\cdot$cm for $x = 0.05$, 13~$\Omega\cdot$cm for $x = 0.1$, and 3~$\Omega\cdot$cm for $x = 0.2$ at 300\,K and reaching a minimum value of $\sim$0.026~$\Omega\cdot$cm at 1100~K for $x = 0.1$. This pronounced reduction in resistivity with Sr content is attributed to enhanced electronic transport associated with Sr substitution and thermally activated hopping transport and is consistent with the changes in $\alpha$ (discussed later). The resistivity values for rare-earth transition metal oxides (shown above) have been compared from literature along with experimental values obtained with simple rare-earth cobaltates (as shown in Fig.~\ref{fig3}(a)), the resistivity of the present multicomponent system remains considerably higher than that of several simple rare-earth transition metal oxides. This behavior may be associated with chemical disorder on the B-site. In contrast to single B-site perovskite, the random distribution of Co, Fe, Mn, Cr, and Ni generates a range of local
electronic environment. Because these cations have different redox potentials, electronic configurations,
and ionic sizes, their random arrangement produces site-energy fluctuations. These effects can partially
localize the charge carriers and increase the activation energy required for hopping. Consequently the
carrier mobility is reduced and the overall resistivity is increased. Besides, as $\mathrm{LaCrO}_{3}$ is an electrical
insulator with large electrical resistivity, an equimolar amount of Cr might be detrimental for the electrical
conduction~\cite{sen2024high}.\\
The temperature dependence of resistivity, $\rho(T)$, is consistent with small-polaron hopping conduction (Fig.~\ref{fig3}(b)), which is commonly observed in cobaltate perovskites with strong electron--phonon coupling and partially localized 3-$d$ carriers~\cite{bonet2016high, benedict2016systematic}:
\begin{equation}
\rho(T) = \rho_0 T \exp\left(\frac{E_\rho}{k_{\mathrm{B}}T}\right),
\end{equation}
where $\rho_0$ is the pre-exponential factor, $E_\rho$ is the resistivity activation energy, and $k_{\mathrm{B}}$ is the Boltzmann constant. The activation energy ($E_\rho$) obtained for each sample is presented in Table~\ref{tab:thermoelectric}.\\ The resistivity activation energy decreases progressively from 0.379 (\textit{x}=0.00) to 0.219 for (\textit{x}=0.2). This suggests that the charge transport becomes easier with increasing Sr substitution. The decrease in E$_\rho$ is accompanied by a strong increase in the electrical conductivity, which further suggests that the Sr doping reduces the energy barrier for polaronic charge transport.
Figure~\ref{fig3}(c) shows the temperature dependence of the Seebeck coefficient ($\alpha$) for La$_{1-x}$Sr$_x$(CoFeMnCrNi)O$_3$. All compositions exhibit positive $\alpha$ throughout the investigated temperature range, confirming dominant hole-like transport. The positive thermopower is consistent with the substitution of Sr$^{2+}$ at the La$^{3+}$ site, which requires charge compensation through changes in the transition-metal valence states and/or oxygen non-stoichiometry. All samples show positive $\alpha$ values throughout the temperature range, confirming dominant p-type conduction. For all compositions, $\alpha$ decreases with increasing temperature, which is consistent with electrical transport by small-polaron
hopping.~\cite{jazandari2026unifying, zhang2024tuning}. At 300~K, $\alpha$ decreases from 182~$\mu$V~K$^{-1}$ for $x = 0.0$ to 157~$\mu$V~K$^{-1}$ for $x = 0.1$ and 94~$\mu$V~K$^{-1}$ for $x = 0.2$, indicating that Sr substitution increases the carrier concentration. Notably, the high-entropy composition La$_{0.9}$Sr$_{0.1}$(CoFeMnCrNi)O$_3$ showed a large value of $\alpha$ compared to simple La$_{0.9}$Sr$_{0.1}$CoO$_3$ (as shown in Fig.~\ref{fig3}(c)), however, as per literature data: La$_{0.9}$Sr$_{0.1}$FeO$_3$ showed $\alpha$ of 487~$\mu$V~K$^{-1}$ at 300~K\cite{wang2010influence}, 25~$\mu$V~K$^{-1}$ at 350~K for La$_{0.9}$Sr$_{0.1}$MnO$_3$ \cite{kuo1990oxidation}, 210~$\mu$V~K$^{-1}$ at 300~K for La$_{0.9}$Sr$_{0.1}$CrO$_3$\cite{stakkestad1996investigation}. The measured $\alpha$ of the five-cation composition is therefore not a simple average of single-cation based system, rather it reflects which of the constituent redox couples is active at the Fermi level and the degeneracy of the associated spin and orbital states~\cite{heremans2008enhancement}.\\
Since the temperature dependent electrical resistivity is described by thermally activated small-polaron hopping, the Seebeck coefficient has been interpreted within a localized-carrier framework. In the high-temperature small-polaron regime, the thermopower can be approximated by the modified Heikes relation\cite{mott2012electronic, koshibae2000thermopower},
\begin{equation}
\alpha(T)=\alpha_{\mathrm{Heikes}}+\frac{E_S}{eT},
\end{equation}
where $\alpha_{\mathrm{Heikes}}$ represents the temperature-independent entropy contribution associated with the configurational, spin, and orbital degrees of freedom of the localized carriers, and $E_S$ is the thermopower activation energy. Accordingly, Fig.~\ref{fig3}(d) presents $\alpha$ as a function of $1000/T$. The data exhibit reasonably good linearity in the high-temperature region between 600\,K and 1100~K, supporting the small-polaron thermopower in this temperature interval. Linear fitting of the high-temperature Seebeck values yields $E_S$ values of approximately 0.077, 0.066, 0.057, and 0.038~eV for T03a, T03b, T03c, and T03d, respectively. A systematic reduction in $E_S$ with increasing Sr content is therefore observed and may be attributed to the energetic asymmetry associated with thermally activated carrier transfer. This behavior is consistent with the simultaneous decrease in electrical resistivity and suggests that Sr incorporation facilitates the transfer of localized hole-like carriers between neighboring transition-metal sites.

The Seebeck coefficient activation energies are substantially smaller than the activation energies derived from electrical resistivity. The corresponding resistivity activation energies, $E_{\rho}$, are approximately 0.379, 0.325, 0.265, and 0.219~eV for T03a--T03d, respectively. Thus, $E_S<E_{\rho}$ for all compositions. Such a difference is expected for small-polaron transport because the electrical resistivity includes an additional energetic contribution associated with lattice-assisted carrier hopping, whereas the Seebeck coefficient is primarily governed by the entropy and energetic asymmetry associated with carrier occupation.

The difference between the two activation energies may be expressed approximately as
\begin{equation}
W_H \simeq E_{\rho}-E_S,
\end{equation}
where $W_H$ represents an effective hopping-related energy contribution\cite{mott2012electronic}. The estimated $W_H$ decreases from approximately 0.302~eV for T03a to 0.259, 0.208, and 0.181~eV for T03b, T03c, and T03d, respectively. The systematic decrease in $W_H$ provides further evidence that Sr substitution progressively reduces the energetic barrier for polaron transfer. This interpretation is consistent with the strong decrease in resistivity observed across the high entropy samples.

The high-temperature intercept, $\alpha_{\mathrm{Heikes}}$, represents the entropy contribution to Seebeck coefficient in the localized-carrier limit. Its magnitude is influenced by the relative populations and degeneracies of the accessible transition-metal valence and spin states. In the present multication B-site sublattice, Co, Fe, Cr, Mn, and
Ni generate a chemically heterogeneous set of local electronic environments. These cations may adopt
different oxidation and spin states, resulting in multiple sites with different energies and carrier
configurations. Substitution of $\mathrm{Sr}^{2+}$ for $\mathrm{La}^{3+}$ modifies the charge-balance conditions and may change the
relative concentrations of the transition-metal valence states, and oxygen defects. Consequently, Sr
incorporation alters both the entropy contribution to the thermopower, reflected by $\alpha_{\mathrm{Heikes}}$ and the activation
energy $E_S$ associated with thermally activated carrier transfer. The power factor for La$_{1-x}$Sr$_x$(CoFeMnCrNi)O$_3$ ($0 \leq x \leq 0.2$) is shown in Fig.~S2.
\\
\begin{figure*}
\centering
  \includegraphics[width=0.7\linewidth]{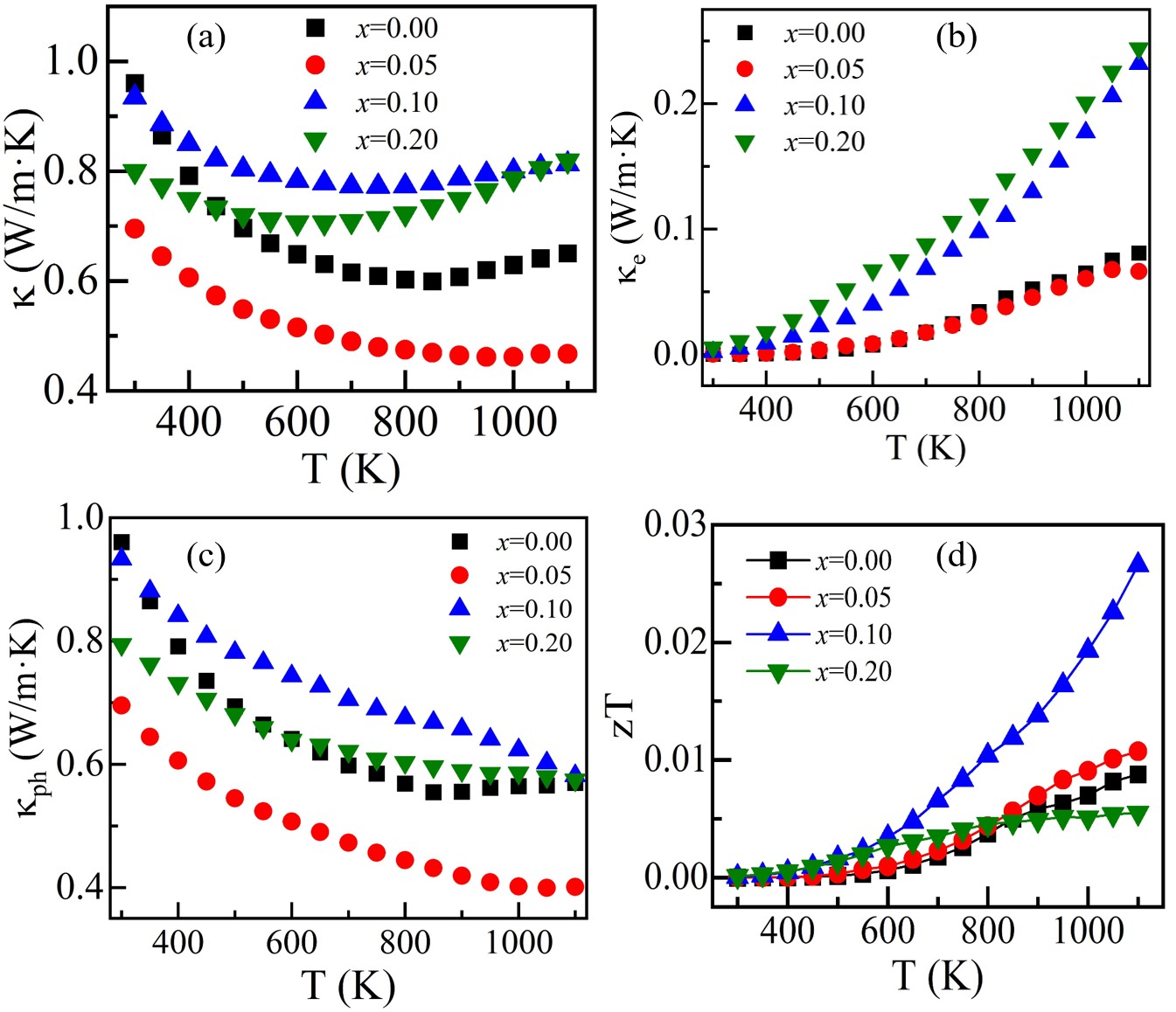}
  \caption{Temperature dependence of (a) total thermal conductivity ($\kappa$), (b) electronic thermal conductivity ($\kappa_e$), (c) phonon thermal conductivity ($\kappa_{ph}$), and (d) thermoelectric figure of merit ($zT$) for 
La$_{1-x}$Sr$_x$(CoFeMnCrNi)O$_3$ ($0 \leq x \leq 0.2$).}
  \label{fig4}
\end{figure*}
Fig.~\ref{fig4}(a) shows the total thermal conductivity ($\kappa$) as a function of temperature (T) for La$_{1-x}$Sr$_x$(CoFeMnCrNi)O$_3$ ($0 \leq x \leq 0.2$). The $\kappa$ values are significantly lower than those of simple perovskite oxides, which is a direct consequence of the multication at B-site in high-entropy perovskites. The $\kappa$ value obtained at 300~K for La$_{1-x}$Sr$_x$(CoFeMnCrNi)O$_3$ with $x = 0.00$ is $\sim$0.95 W/(m·K). For the sample with \textit{x}=0.05, $\kappa$ decreases drastically $(\sim 0.7\,\mathrm{W\,m^{-1}\,K^{-1}}\text{ at }300\,\mathrm{K})$ as compared to the parent compound. This reduction may be related with the heterogeneous microstructure present in this sample (as seen from the SEM image). Because electrical transport in these compositions is governed by small-polaron hopping, the $\kappa_e$ was estimated using the small-polaron form of the Wiedemann-Franz relation, in which the Lorenz number is governed by the hopping activation energy:\cite{wood1985thermal}
\begin{equation}
\kappa_e = L\,\sigma\,T,
\qquad
L = \left(\frac{E_\rho}{eT}\right)^{2},
\label{eq:lorenz}
\end{equation}
so that
\begin{equation}
\kappa_e = \frac{E_\rho^{2}\,\sigma}{e^{2}T} = \frac{E_\rho^{2}}{e^{2}\rho T},
\label{eq:kappae}
\end{equation}
where $E_\rho$ is the resistivity activation energy obtained from the linear fit of $\ln(\rho/T)$ versus $1000/T$ (Table~\ref{tab:thermoelectric}), $e$ is the elementary charge, and $\rho = 1/\sigma$ is the measured electrical resistivity. The Lorenz numbers obtained from Eq.~(\ref{eq:lorenz}) are considerably larger than $L_0$ (for example, $L \approx 5.8 \times 10^{-8}$~W\,$\Omega$\,K$^{-2}$ at 1100~K for $x = 0.10$), reflecting the fact that each hop transports an energy of the order
of $E_\rho$ in addition to the carrier charge. The resulting $\kappa_e(T)$ is shown in Fig.~\ref{fig4}(b). At 300~K, $\kappa_e$ is negligible for all compositions ($\lesssim 10^{-3}$~W/(m·K), i.e.\ below 1\% of $\kappa$), which is a direct consequence of the very low carrier mobility in the localized-carrier regime. With increasing temperature, $\kappa_e$ rises steeply
because the exponential activation of $\sigma$ outweighs the $1/T^{2}$ prefactor in Eq.~(\ref{eq:kappae}); this behaviour persists as long as $E_\rho > 2k_BT$, which holds throughout the measured range for the present activation energies. At 1100~K, $\kappa_e$ reaches $\sim$0.25~W/(m·K) for $x = 0.10$ and $x = 0.20$, whereas it remains below $\sim$0.08~W/(m·K) for $x = 0.00$ and $x = 0.05$, consistent with the systematic reduction of $\rho$ and $E_\rho$ with Sr substitution. Even for the most conducting composition, $\kappa_e$ accounts for less than one-third of $\kappa$ at 1100~K, confirming that the total thermal
conductivity remains phonon dominated across the entire series.\\
Consequently, $\kappa_{ph}$ [Fig.~\ref{fig4}(c)] closely follows $\kappa$ at low temperature and decreases monotonically with increasing temperature, as expected for Umklapp-limited transport superimposed on strong point-defect scattering. The low absolute magnitude of $\kappa_{ph}$
(0.4--0.95~W/(m·K)) confirms that the mass fluctuation, strain-field fluctuation, and local bond distortion introduced by five different cations at the B-site scatter phonons strongly and drive $\kappa_{ph}$ towards lower value. The mild flattening of $\kappa$ above $\sim$800~K for $x = 0.10$ and $x = 0.20$ arises from the growing $\kappa_e$ compensating the continued decrease in $\kappa_{ph}$.\\
The resulting figure of merit is presented in Fig.~\ref{fig4}(d). The $zT$ increases monotonically with temperature for all compositions, reaching a maximum of $\sim$0.027 at 1100~K for $x = 0.10$, which represents the best compromise between the reduction in $\rho$ and the accompanying decrease in $\alpha$ with Sr substitution.\\
\subsection{Non-Equimolar High Entropy Oxides}
The equimolar B-site composition with optimal Sr doping yields a maximum $zT \sim 0.027$ at 1100~K, despite a $\kappa_{ph}$ below 1 W/(m·K), showing that strong phonon scattering alone is insufficient to maximize TE performance. In this view, additional compositions with non-equimolar B-site distributions and selected multi-rare-earth A-site substitutions were synthesized, as described in the experimental section.
\begin{figure*}
\centering
  \includegraphics[width=0.7\linewidth]{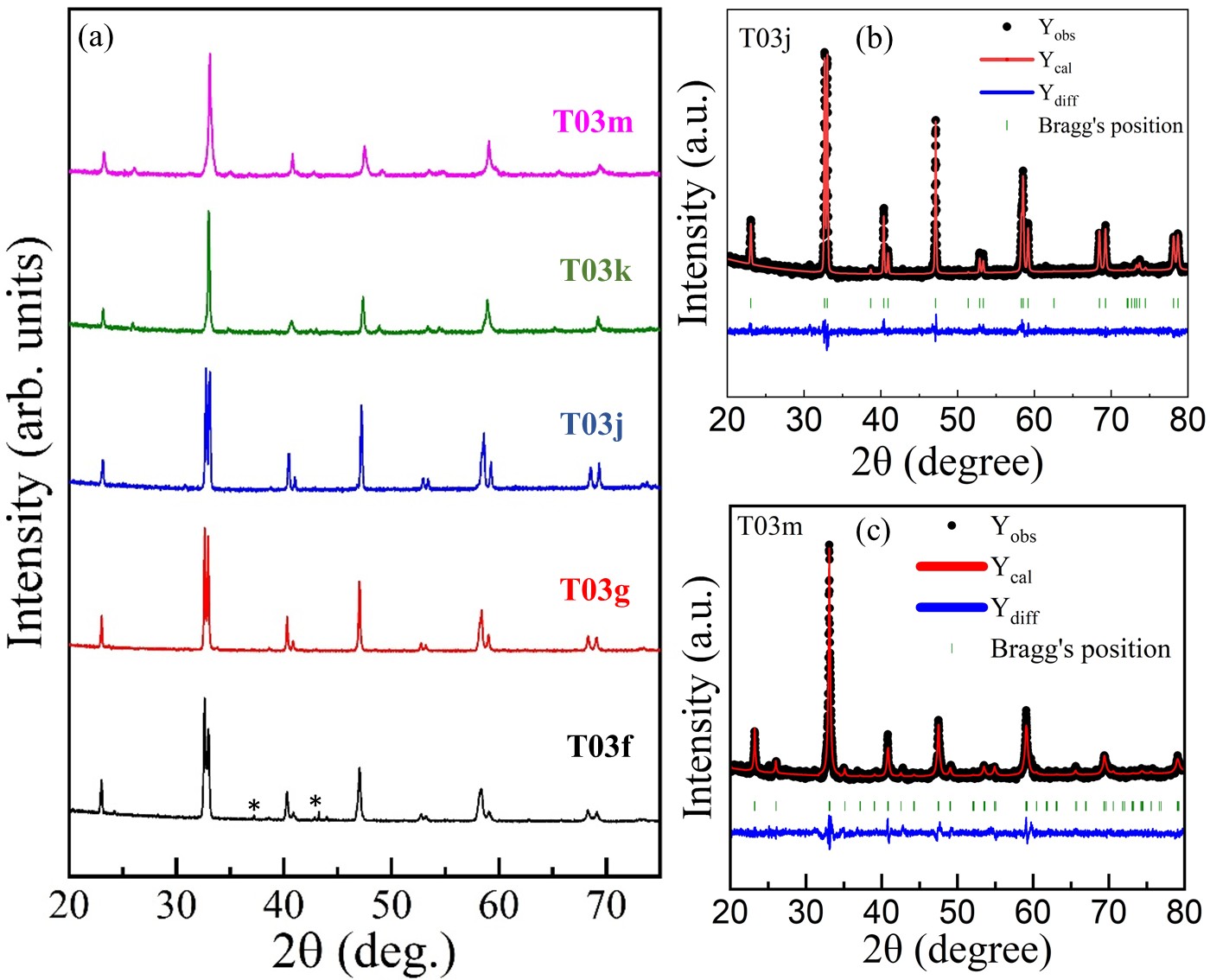}
  \caption{(a) X-ray diffraction (XRD) pattern for different non-equimolar high entropy samples. Rietveld refinement pattern for (b) T03j and (c) T03m is shown.}
  \label{fig5}
\end{figure*}
\subsubsection{X-ray Diffraction}
The XRD patterns of these non-equimolar compositions are shown in Fig.~\ref{fig5}(a), which confirms the formation of perovskite structures. It is noted that varying the concentrations of elements at the A- and B-sites results in two distinct XRD patterns which were ascribed with two different space groups, namely $R\bar{3}c$ and $Pnma$, which is consistent with the tolerance factors calculated for each sample, as shown in Table~\ref{tab:structural_parameters}. Only T03f shows a few obvious weak secondary reflections, which is consistent with its size-disorder parameter of 5.42\% (Table~\ref{tab:structural_parameters}), the largest in the series and at the limit of the empirical $\delta \leq$5\% single phase criterion~\cite{zhang2008solid}. The XRD patterns of all the samples were analyzed using Rietveld refinement, and the lattice parameters obtained are listed in Table~\ref{tab:refinement} and the refinement patterns for T03j ($R\bar{3}c$) and T03m ($Pnma$) are shown in Fig.~\ref{fig5}(b,c) and the refinement patterns for the entire series is presented in Fig.S3.\\ 
\subsubsection{High-resolution Transmision Electron Microscopy}
\begin{figure*}
\centering
  \includegraphics[width=0.85\linewidth]{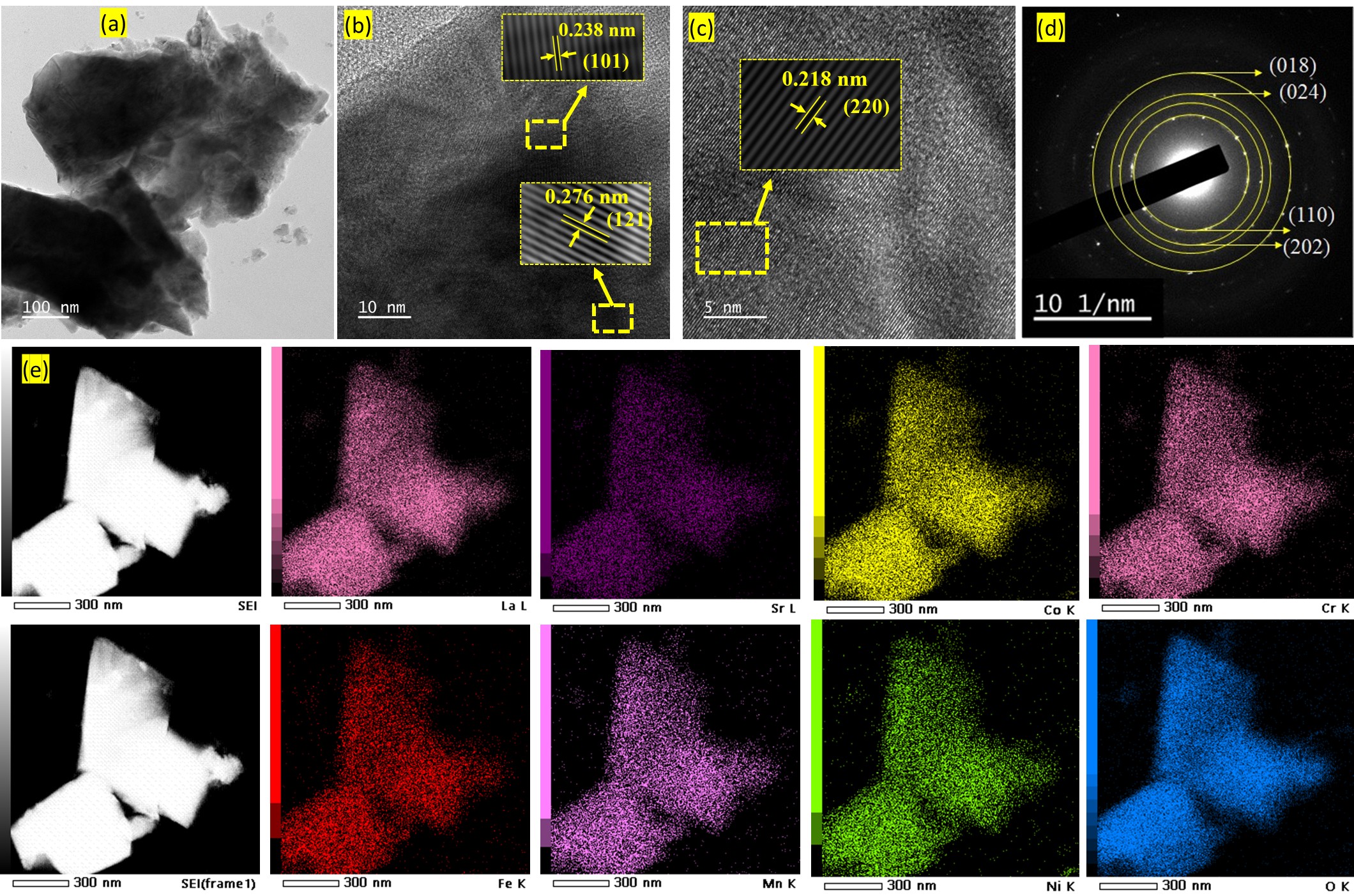}
  \caption{(a) TEM image, (b,c) HRTEM images showing resolved lattice fringes, (d) SAED pattern indexed to the perovskite phase for T03j sample and (e) corresponding elemental maps of La, Sr, Co, Cr, Fe, Mn, Ni, and O, demonstrating uniform elemental distribution in the sample.}
  \label{fig6}
\end{figure*}

Figure~\ref{fig6} presents the TEM, HRTEM, SAED, and elemental mapping analyses of the T03j sample. The low-magnification TEM image (Fig.~\ref{fig6}a) reveals agglomerated particles with irregular morphology, consistent with SEM observations (Fig.~S4). The HRTEM images (Fig.~\ref{fig6}b,c) exhibit well-defined lattice fringes with interplanar spacings of 0.276, 0.238, and 0.218 nm, corresponding to the (121), (101), and (220) crystallographic planes, respectively, confirming the highly crystalline nature of the material. The selected-area electron diffraction (SAED) pattern (Fig.~\ref{fig6}d) displays distinct diffraction rings indexed to the (110), (018), (024), and (202) planes, further validating the polycrystalline perovskite structure. Elemental mapping results (Fig.~\ref{fig6}e) demonstrate a homogeneous spatial distribution of La, Sr, Co, Cr, Fe, Mn, Ni, and O throughout the analyzed region without any noticeable elemental segregation. These observations confirm the good chemical homogeneity of the samples at the nanoscale.\\
\subsubsection{X-ray photoelectron spectroscopy}
\begin{figure*}
\centering
  \includegraphics[width=0.8\linewidth]{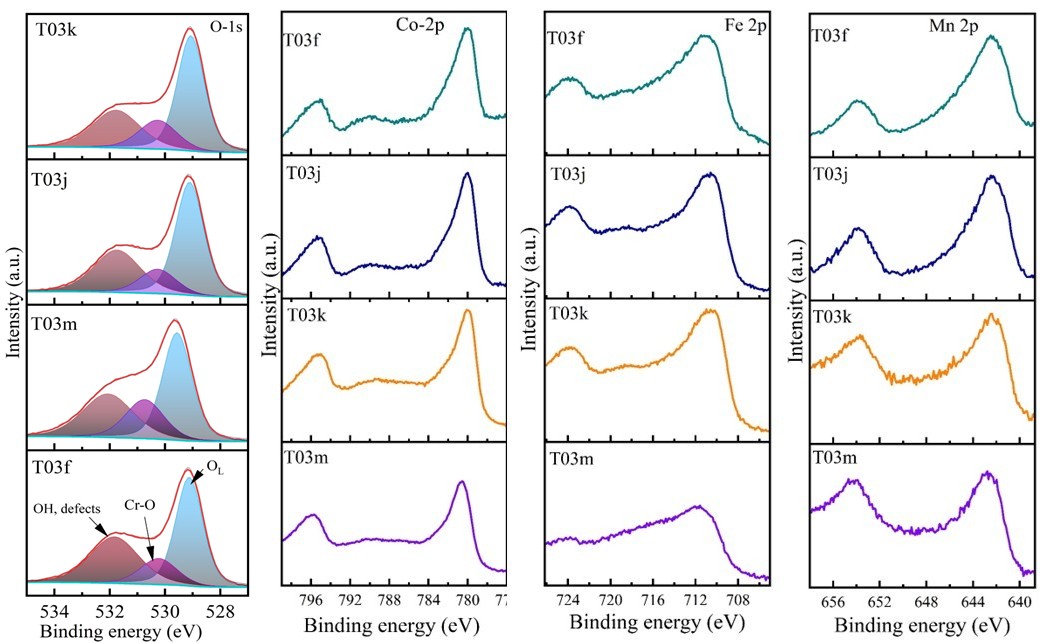}
  \caption{Core-level X-ray photoelectron spectroscopy (XPS) spectra of selected 
non-equimolar high-entropy oxide compositions T03f, T03j, T03k, and T03m: (a) O-1$s$, (b) Co-2$p$, (c) Fe-2$p$, and (d) Mn-2$p$.}
  \label{fig7}
\end{figure*}

The XPS survey spectra (Fig.~S5) of samples T03f, T03j, T03k and T03m further confirm the presence of all constituent elements of the high-entropy perovskite oxide. Besides the characteristic peaks of La, Cr, Mn, Fe, Co, Ni and O, the additional rare-earth elements introduced in these compositions are also detected. Owing to the multicomponent nature of these compositions, significant overlap among several core-level spectra is observed. In particular, La-3$d$ overlaps with Ni-2$p$, Co-2$p$ overlaps with Ni and Fe Auger features, Fe-2$p$ overlaps with Co Auger peaks, while Mn-2$p$ is partially interfered by Ni Auger signals, as mentioned earlier. For T03k and T03m, further overlap involving Sm, Nd, Pr and Dy core levels makes reliable quantification of most transition-metal spectra impractical. Consequently, quantitative determination of the oxidation states of individual transition-metal cations was not attempted.\\
The core-level O~1$s$ spectra (Fig.~\ref{fig7}(a)) provide valuable information regarding the surface oxygen chemistry. The spectra of T03f and T03j can be satisfactorily resolved into lattice oxygen (O$_L$), defect-related oxygen (O--H, defects), and a weak intermediate component centred near 530.2--530.3~eV, which may originate from Cr--O surface species~\cite{easton2025critical}. The O$_L$ appears at approximately 529.2$\pm0.2$~eV, whereas the defect-related oxygen component occurs at 531.5~eV. The relative contribution of defect oxygen remains appreciable, with $\mathrm{O{-}C/O{-}H}$, defect area ratios of approximately 0.76 and 0.61 for T03f and T03j, respectively, indicating a high concentration of oxygen-defect and surface hydroxyl species. The intermediate oxygen component contributes about 0.33 and 0.28 of the lattice oxygen area for T03f and T03j, respectively, suggesting the possible presence of surface chromium oxide species. For T03k and T03m, the O~1$s$ spectra exhibit a similar three-component nature consisting of lattice oxygen, defect-related oxygen and an intermediate oxygen species. The defect-related oxygen contribution remains significant, with relative area ratios of approximately 0.56 for T03k and 0.69 for T03m, while the intermediate component accounts for approximately 0.35 and 0.47 of the lattice oxygen area, respectively. These observations indicate that oxygen-defect formation continues to be an important charge-compensation mechanism in these highly substituted compositions.\\
Although the transition-metal core-level spectra exhibit characteristic Co~2$p$, Fe~2$p$, Mn~2$p$ and Cr~2$p$ features (Fig.~\ref{fig7}), extensive peak overlap arising from Auger transitions and neighbouring rare-earth core levels prevents reliable quantitative fitting of individual oxidation states, as mentioned earlier. Nevertheless, the observed binding energies remain consistent with transition metals existing predominantly in higher oxidation states typical of perovskite oxides, with possible contributions from subtle variations in transition-metal valence states that cannot be resolved unambiguously by the present XPS measurements.\\
\begin{table}[htbp]
\centering
\caption{Refinement parameters viz. space group and lattice parameters ($a$, $b$, $c$), goodness of fit ($\chi^2$) and relative density for all the high-entropy samples.}
\label{tab:refinement}
\begin{tabular}{ccccc}
\hline
Sample & Space group & $a$ (\AA) & $b$ (\AA) & $c$ (\AA) \\
\hline
T03a & $Pnma$ & 5.518(9) & 7.752(3) & 5.481(4)\\
T03b & $Pnma$ & 5.521(7) & 7.769(9) & 5.491(6) \\
T03c & $R\bar{3}c$ & 5.485(3) & 5.485(3) & 13.255(3) \\
T03d & $R\bar{3}c$ & 5.500(5) & 5.500(5) & 13.401(3)\\
T03f & $R\bar{3}c$ & 5.504(8) & 5.504(8) & 13.284(0) \\
T03g & $R\bar{3}c$ & 5.491(4) & 5.491(4) & 13.260(5) \\
T03j & $R\bar{3}c$ & 5.477(4) & 5.477(4) & 13.216(7) \\
T03k & $Pnma$ & 5.453(0) & 7.663(1) & 5.411(6) \\
T03m & $Pnma$ & 5.432(5) & 7.717(8) & 5.505(5)\\
\hline
\end{tabular}
\end{table}
\begin{table*}[htbp]
\centering
\caption{Thermoelectric parameters ($\alpha$, $\sigma$ and $\kappa$) at 300~K and resistivity activation energy (E$_{\rho}$) obtained from ln($\rho/T$) vs 1000/T plot, Seebeck activation energy (E$_{\alpha}$) obtained from $\alpha$ vs 1000/T and E$_{\rho}$-E$_{\alpha}$ for all the high-entropy samples.}
\label{tab:thermoelectric}
\begin{tabular}{ccccccccc}
\hline
Sample & $\alpha$ & $\sigma$ & $\kappa_{mes.}$ & $\kappa_{dens.}$ & $E_{\rho}$ &$E_{\alpha}$& E$_{\rho}$-E$_{\alpha}$  & $\alpha_{Heikes}$ \\
Sample & ($\mu$V/K) &  (S/cm) &  (W/(m·K)) & (W/(m·K))& (eV) &(eV)&  (eV) & ($\mu$V/K)\\
\hline
T03a & 181 & 0.0008 & 0.96 & 1.34 & 0.379 & 0.077 & 0.302&23.2\\
T03b & 169 & 0.0041 & 0.70 & 1.07 & 0.325 & 0.066&0.259&22.8\\
T03c & 156 & 0.07 & 0.93 & 1.32 & 0.265 & 0.057&0.208&22.4\\
T03d & 95 & 0.3 & 0.80 & 1.08 & 0.219 & 0.0379&0.181&12.4\\
T03f & 50 & 4.16& 0.76 & 1.06 & 0.144 & 0.034&0.110&28.2\\
T03g & 154 & 0.65& 0.94 & 1.34 & 0.21 & 0.0464&0.164&20.8\\
T03j & 220 & 0.44& 0.66 & 0.89 & 0.19 & 0.0466 &0.144&2.5\\
T03k & 148 & 0.71& 0.56 & 0.83 & 0.204 &0.0482 &0.159&17.8\\
T03m & 252 & 0.058& 0.56 & 0.78 &0.26 & 0.0912&0.169&1.3\\
\hline
\end{tabular}
\end{table*}
\subsubsection{Electronic and Thermal Transport}
\begin{figure*}
\centering
  \includegraphics[width=0.8\linewidth]{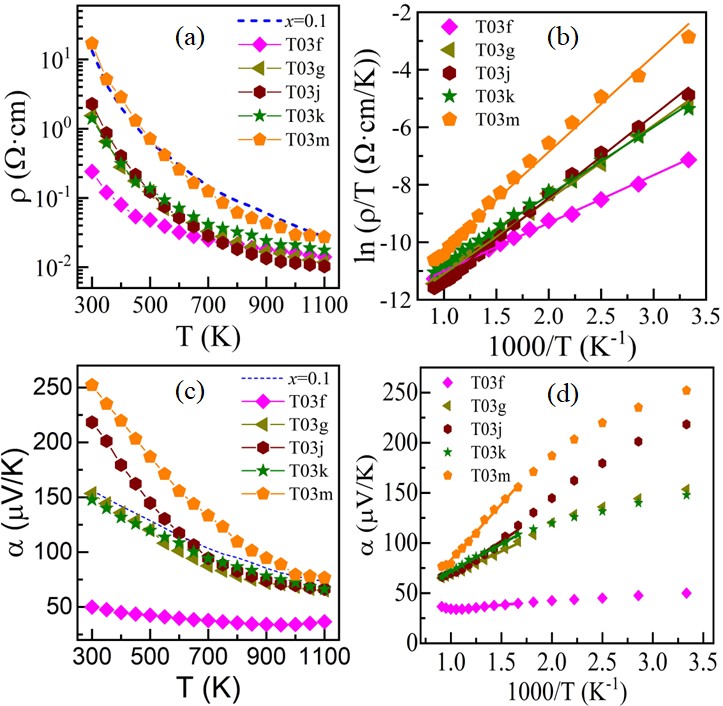}
  \caption{Temperature dependence of (a) electrical resistivity ($\rho$), (b) ln($\rho$/T) vs 1000/T (c) Seebeck coefficient ($\alpha$), and (d) $\alpha$ vs 1000/T for T03f, T03g, T03j, T03k, and T03m. The $\rho$ and $\alpha$ for non-equimolar high entropy samples are compared with La$_{0.9}$Sr$_{0.1}$(CoFeMnCrNi)O$_3$.}
  \label{fig9}
\end{figure*}
\begin{figure}[h]
\centering
  \includegraphics[width=0.8\linewidth]{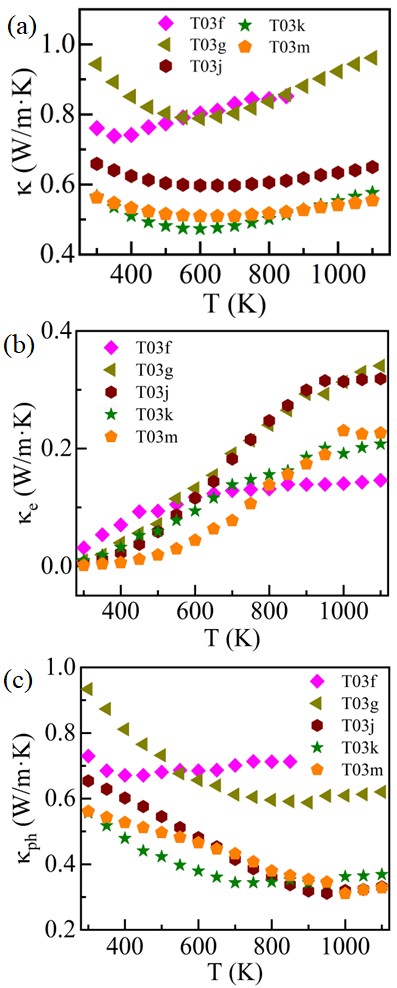}
  \caption{Temperature dependence of (a) total thermal conductivity ($\kappa$), (b) electronic thermal conductivity ($\kappa_e$), and (c) phonon thermal conductivity ($\kappa_{ph}$) for T03f, T03g,
T03j, T03k, and T03m.}
  \label{fig8}
\end{figure}

Electrical resistivity ($\rho$) as a function of temperature for all high-entropy samples is shown in Fig.~\ref{fig9}(a). The resistivity trends are broadly consistent with the Seebeck-coefficient behavior: samples with the highest $\alpha$ also tend to show higher resistivity, reflecting the strong interdependence of thermoelectric transport parameters~\cite{walia2013transition}. It is important to note that two compositions T03f and T03g contain same five cations on the B-site and they share similar A-site composition; they differ only in which cations are present in excess. Their room temperature $\sigma$ differ by a factor of 6.4 (4.16 vs 0.65 S/cm) and their $\alpha$ by a factor of 3.1 (50 vs 154 $\mu$V/K). It follows directly from cation chemistry: T03f is Ni-rich and Ni$^{3+}$ in a LaNiO$_3$-like octahedral environment supplies high concentration of itinerant carriers, collapsing both the hopping barrier ($E_\rho$=0.144 eV) and $\alpha$; T03g is Co/Fe rich, retaining the localised, spin-degenerate Co$^{3+}$/Co$^{4+}$ manifold that sustains $\alpha$. The corresponding $E_\rho$ =0.144 eV for T03f against 0.21 eV for T03g at identical configurational entropy shows that the polaron barrier is purely chemical. These results again indicate that simply maximizing configurational entropy is insufficient; instead, the entropy level and cation distribution must be tuned to maintain adequate electronic transport. Importantly, several non-equimolar high-entropy compositions show resistivity lower than that of the equimolar analog, demonstrating that compositional optimization can partially recover electronic transport while retaining strong phonon scattering. The electrical resistivity of T03j is 2.3~$\Omega\cdot$cm at 300~K and decreases to 0.01~$\Omega\cdot$cm at 1100~K. Such a decrease in electrical resistivity may be attributed to the combined effects of Sr-induced hole doping, and thermally activated transport. Arrhenius plot for these non-equimolar samples are shown in Fig.~\ref{fig9}(b). The good linearity of the ln($\rho$/T) versus 1000/T plots indicates that electrical transport is governed predominantly by thermally activated adiabatic small-polaron hopping over the investigated temperature range. The activation energy decreases from 0.38 eV for T03a to 0.22\,eV for T03d (with Sr doping) and it changes effectively with different composition and their concentration in each sample, indicating that compositional engineering tunes the hopping barrier for charge transport.\\
Fig.~\ref{fig9}(c) presents the temperature-dependent Seebeck coefficient of the non-equimolar high-entropy compositions. All samples remain p-type, and $\alpha$ decreases with increasing temperature, consistent with semiconducting transport. The Seebeck coefficient is found to be highly sensitive to the concentration and distribution of transition-metal cations at the B-site, varying from $\sim$220 to $\sim$50~$\mu$V~K$^{-1}$ at 300~K depending on the dominant element in the composition. Compositions with higher Co content generally exhibit larger $\alpha$ values, as Co possesses a combination of mixed valence, spin-state degeneracy, narrow-band transport, and strong electron correlation~\cite{asai1998two,smith2012evolution}. The different spin states create additional spin and orbital entropy, thereby improving the Seebeck coefficient, since $\alpha$ is related to the entropy transported per charge carrier. The Seebeck coefficient for the Co-rich sample T03j is $\sim$220~$\mu$V~K$^{-1}$ at 300~K. Further, the samples with higher Ni content show smaller $\alpha$ values ($\sim$50~$\mu$V~K$^{-1}$), which may be due to the higher carrier concentration generally observed in LaNiO$_3$-based systems~\cite{vulchev2012improving}. The high-entropy sample with higher Cr content also shows a considerable $\alpha$ value, as Cr$^{3+}$ typically suppresses excessive carrier mobility and reduces carrier concentration, thereby improving $\alpha$. After achieving considerable $\alpha$ through B-site substitutions, the A-site was further populated with five different cations, and the corresponding temperature-dependent $\alpha$ values are shown in Fig.~\ref{fig9}(c). T03m shows the highest $\alpha$ value of $\sim$250~$\mu$V~K$^{-1}$ at 300~K. This enhancement may be attributed to the increased configurational entropy, as well as changes in the average ionic radius at the A-site, which modify the local environment of the BO$_6$ octahedra and hence affect the transport properties, as electrical transport in these compositions is governed by small-polaron hopping. For thermally activated small polarons, the Seebeck coefficient can be approximated using Heikes formula. The linear fit in high temperature region suggest the small-poaron hopping in non-equimolar high entropy systems. 
As a result of the simultaneous optimization of electrical resistivity and Seebeck coefficient, the power factor ($\alpha^2\sigma$), where $\sigma = 1/\rho$, was calculated from the electrical resistivity. The power factor increases with temperature for most optimized compositions (Fig.~S6), mainly due to the significant reduction in electrical resistivity at high temperature. Among all samples, T03j exhibits the highest power factor, reaching approximately 40--43~$\mu$W~m$^{-1}$~K$^{-2}$ near 1000--1100~K. This superior power factor arises from the optimum combination of a reasonably high Seebeck coefficient and enhanced electrical conductivity. T03g also shows a relatively high power factor, reaching approximately 35--36~$\mu$W~m$^{-1}$~K$^{-2}$ at 1100~K. Although T03m shows the highest Seebeck coefficient, its higher resistivity limit its power factor. Conversely, T03f has low resistivity but also a very low Seebeck coefficient, resulting in the lowest power factor among the investigated compositions. This suggests that the power factor of samples with non-equimolar B-site concentrations is higher than that of the equimolar composition.\\
Fig.~\ref{fig8}(a-c) shows variation of $\kappa$, $\kappa_e$, and $\kappa_{ph}$ as a function of temperature for different high-entropy oxide compositions. The thermal conductivity remains strongly suppressed throughout the composition and temperature range. At room temperature, $\kappa$ is $\sim$0.95 W/(m·K) for 
La$_{0.9}$Sr$_{0.1}$(Co$_{0.2}$Fe$_{0.2}$Mn$_{0.2}$Cr$_{0.2}$Ni$_{0.2}$)O$_3$, 
decreases to $\sim$0.76 W/(m·K) for La$_{0.9}$Sr$_{0.1}$(Co$_{0.15}$Fe$_{0.15}$Mn$_{0.15}$Cr$_{0.15}$Ni$_{0.4}$)O$_3$, 
and is $\sim$0.95 W/(m·K) for La$_{0.9}$Sr$_{0.1}$(Co$_{0.3}$Fe$_{0.3}$Ni$_{0.2}$Cr$_{0.1}$Mn$_{0.1}$)O$_3$. 
A further reduction to $\sim$0.56 W/(m·K) is observed for the compositions containing multiple substitutions at both A- and B-sites. As the relative densities for all the samples are between 75\%-81\%, Maxwell-Eucken relation~\cite{francl1954thermal} for estimating the thermal conductivity of 100\% dense materials is used as $\kappa_{dens.}$=$\kappa_{meas.}(\frac{1+P/2}{1-P})$; where P is the porosity present in the material. The $\kappa_{dens.}$ for all the compositions have been presented in Table~\ref{tab:thermoelectric}. Since $\rho$ also depends on the porosity in a similar way as $\kappa$, so it may not effect overall zT.\\
As can be seen from Fig.~\ref{fig8}(b), $\kappa_e$ (estimated using small-polaronic model as discussed earlier) is small at lower temperature and increases with increasing temperature and is attributed to increase in $\sigma$. Temperature-dependent ($\kappa_{ph}$) shown in Fig.~\ref{fig8}(c) decreases significantly with increasing temperature. This may be attributed to enhanced phonon scattering with rise in temperature. The variation of $\kappa$ across this series is well described by point-defect phonon scattering. At fixed A-site composition (La$_{0.9}$Sr$_{0.1}$), the room-temperature $\kappa$ decreases as the B-site mass-variance parameter
increases: $\kappa = 0.94$, 0.93, 0.76 and 0.66~W/(m·K) for $\Gamma_{M}(\mathrm{B}) = 1.55$, 2.10, 1.94 and $2.88 \times 10^{-3}$ (T03g, T03c, T03f, T03j), with the departure of T03f accounted for by its anomalously large size-disorder parameter of 5.42\%, which contributes an
additional strain-field term, coupled to the presence of small amounts of impurity phases that may also act as phonons scattering centers. The further reduction to 0.56~W/(m·K) in T03k and T03m follows the same rule applied to the A-site: substituting the equiatomic La--Nd--Pr--Dy--Sm mixture raises $\Gamma_{M}(\mathrm{A})$ from 13.2 to $19.3 \times 10^{-3}$, adding a second mass-fluctuation channel, and simultaneously increases octahedral
tilting. Both are mass- and geometry-based mechanisms in the Klemens sense; neither depends on the number of species \emph{per se}. It is worth noting that the largest single contribution to $\Gamma_{M}(\mathrm{A})$ in every Sr-containing sample comes from the Sr--La mass difference itself, so that the nominally ``low-entropy'' doped compositions already carry substantial A-site
point-defect scattering.\\
\begin{figure}
\centering
  \includegraphics[width=0.8\linewidth]{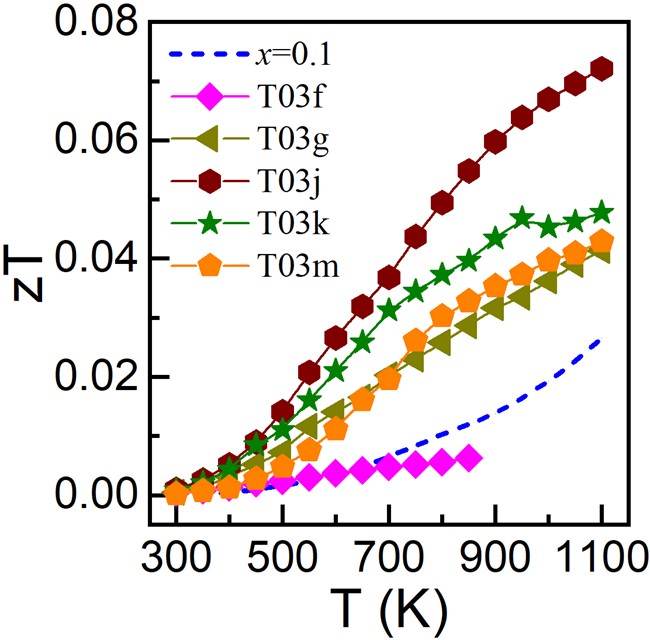}
  \caption{Figure of merit ($zT$) for T03f, T03g, T03j, T03k, and T03m, compared with La$_{0.9}$Sr$_{0.1}$(CoFeMnCrNi)O$_3$.}
  \label{fig10}     
\end{figure}
The dimensionless figure of merit ($zT$) as a function of temperature is shown in Fig.~\ref{fig10}. It follows a trend similar to the power factor and increases with temperature for all compositions. T03j (La$_{0.9}$Sr$_{0.1}$Co$_{0.4}$Cr$_{0.3}$Fe$_{0.1}$Ni$_{0.1}$Mn$_{0.1}$O$_3$) exhibits the highest $zT$, reaching $\sim$0.072 at 1100~K. This value is among the promising values reported for p-type LaCoO$_3$-based and high-entropy perovskite oxide thermoelectrics in this temperature range, particularly considering the very low thermal conductivity and air-stable oxide composition. T03k and T03m show intermediate $zT$ values of around 0.04--0.05 at high temperature, while T03f shows the lowest $zT$. The enhanced $zT$ of T03j originates from a specific cation selection rather than from a disorder regime. Its B-site is 40\% Co, supplying the mixed-valence, spin-degenerate manifold responsible for
$\alpha = 220~\mu$V~K$^{-1}$, and 30\% Cr$^{3+}$, which is redox-inert and therefore limits the carrier concentration without contributing localised states at the Fermi level, while the remaining 30\% distributed over Ni, Fe and Mn is sufficient to keep $\Gamma_{M}(\mathrm{B})$ at the highest value in the series and $\kappa$ correspondingly low. These results indicate that the thermoelectric performance of high entropy perovskite oxide is not governed solely by high configurational entropy or high Seebeck coefficient, but rather by an optimized balance among Seebeck coefficient, electrical resistivity and lattice disorder.\\
\section{Conclusion}
In summary, compositional-engineering at B-site in high-entropy rare-earth cobaltate system provides a viable strategy for improving the thermoelectric performance. In La$_{1-x}$Sr$_x$(CoFeMnCrNi)O$_3$ system, Sr substitution increases the hole concentration and reduces resistivity, while the multication B-site configuration strongly suppresses thermal conductivity through disorder-induced phonon scattering. However, the equimolar high-entropy composition does not yield the best performance, indicating that maximum configurational entropy may be insufficient. Instead, non-equimolar B-site distributions and additional A-site complexity provide a more favorable balance among the Seebeck coefficient, electrical conductivity, and thermal conductivity. Among the investigated samples, La$_{0.9}$Sr$_{0.1}$Co$_{0.4}$Cr$_{0.3}$Ni$_{0.1}$Fe$_{0.1}$Mn$_{0.1}$O$_3$ exhibits the best overall thermoelectric response, reaching $zT \sim 0.072$ at 1100~K, 2.7 times the equimolar analogue, by
combining a Co-rich sublattice that sustains a large $\alpha$ with redox-inert
Cr$^{3+}$ that limits carrier concentration while lowering the hopping barrier to
0.19~eV. The present study demonstrates that multicomponent oxide thermoelectrics should therefore be designed by selecting cations for their electronic function and mass contrast, rather than by maximising the entropy of mixing.\\
\section*{CRediT authorship contribution statement}
{\textbf{Jitendra Kumar:} Investigation, Formal analysis, Writing-original draft. \textbf{David B\'erardan:} Methodology, Writing-review \&editing. \textbf{Diana Dragoe:} Investigation (XPS), Writing-review \& editing. \textbf{Nita Dragoe:} Writing-review \& editing. \textbf{Ashutosh Kumar:} Conceptualization, Formal analysis, Supervision, Writing-original draft, Writing-review \& editing.}

\section*{Declaration of Competing Interest}
The authors declare that they have no known competing financial interests or personal relationships that could have appeared to influence the work reported in this paper.

\section*{Data availability}
Data will be made available on reasonable request from the corresponding author.

\section*{Acknowledgements}
J. Kumar acknowledge Ministry of Education India for PhD fellowship. J. Kumar and A. Kumar also thank the CIF, IIT Bhilai for the experimental facilities.\\
\bibliographystyle{elsarticle-num}
\bibliography{ref}
\end{document}